# Multiphoton-pumped UV-Vis transient absorption spectroscopy of 2D materials: basic concepts and recent applications

Yuri D. Glinka[1,2, a]

[1] The institute of Optics, University of Rochester, Rochester, NY 14627, USA
[2] Institute of Physics, National Academy of Sciences of Ukraine, Kyiv 03028, Ukraine

[a] Electronic mail: yuridglinka@yahoo.com

**Abstract**

2D materials are considered a key element in the development of next-generation electronics (nanoelectronics) due to their extreme thickness in the nanometer range and unique physical properties. The ultrafast dynamics of photoexcited carriers in such materials are strongly influenced by their interfaces, since the thickness of 2D materials is much smaller than the typical depth of light penetration into their bulk counterparts and the mean free path of photoexcited carriers. The resulting collisions of photoexcited carriers with interfacial potential barriers of 2D materials in the presence of a strong laser field significantly alter the overall dynamics of photoexcitation, allowing laser light to be directly absorbed by carriers in the conduction/valence band through the inverse bremsstrahlung mechanism. The corresponding ultrafast carrier dynamics can be monitored using multiphoton-pumped UV-Vis transient absorption spectroscopy. In this review, we discuss the basic concepts and recent applications of this spectroscopy for a variety of 2D materials, including transition-metal dichalcogenide monolayers, topological insulators, and other 2D semiconductor structures.

**Introduction**

Since the discovery of the first single atomic layer material, graphene [1], the dynamics of ultrafast relaxation of photoexcited carriers in such a 2D material and a wide range of its single-layer analogues have been intensively studied using various pump-probe optical methods [2–30]. In general, 2D materials include graphene, boron nitride, transition-metal dichalcogenide monolayers, phosphorene, topological insulators (TI), 2D derivatives of the organic-inorganic hybrid perovskites, and other molecular systems including associated allotropes and compounds [1-30]. 2D materials are very promising for the next generation of electronics (nanoelectronics) due to their extreme thickness and unique physical properties. Specifically, the extreme thickness means that the depth of light penetration into bulk counterparts of 2D materials and the mean free path of photoexcited carriers significantly exceed their thickness [31–36]. This behavior expands the family of layered 2D materials to quasi-2D materials, such as single-layer 3D nanocrystals with a size range of several nanometers [37]. In most cases, 2D and quasi-2D materials and heterostructures based on them (below we will refer to all these structures as 2D materials) are grown or mechanically exfoliated from bulk crystals on transparent or opaque substrates, which are expected to have a strong impact on the dynamics of photoexcited carriers [6, 38–40].

One of the efficient methods for studying ultrafast carrier dynamics in 2D materials is UV-Vis transient absorption (TA) spectroscopy, implemented in transmission or reflection geometry [2-13, 24, 26-30, 37-40]. In the latter case, to switch to the absorption scale, the Kramers-Kronig transformation is necessary [10]. It is generally taken for granted that pumping in all UV-Vis TA experiments typically occurs in the one-photon absorption regime. Thus, this behavior is identical to what occurs in all pump-probe experiments that have used other non-optical probing techniques such as time- and angle-resolved photoemission spectroscopy [TrARPES, also known as two-photon photoemission spectroscopy (2PPES)] [14–23], time-resolved X-ray diffraction (TrXRD) [41,42] and time-resolved conductivity [43]. Although all probing methods reflect the influence of pump-excited carriers on semiconductor properties, only 2PPES and one-photon-pumped TA spectroscopy (1PPTAS) allow one to study the evolution of the electron energy distribution function (EDF) during the relaxation process [14–23, 37]. In addition to elucidating the carrier cooling mechanism, EDF also provides valuable information about the possible states of carriers in semiconductor devices. This unique behavior is due to the fact that photoemission of pump-excited electrons in 2PPES is allowed from any electronic state in the conduction band, similar to the Pauli blocking mechanism in 1PPTAS, which probes all electronic states through carrier population dynamics [Fig. 1(a)].

However, it has recently been shown that TA spectra of TI $Bi_2Se_3$ thin films extend to the entire Vis range and part of the UV range, although the energy of the pump photons lies in the near-IR range [24, 26, 28]. A similar spectroscopic upconversion, even under Vis range pumping, was also observed in TA spectra of single-layer $MAPbBr_3$ nanocrystals [37], monolayers of transition-metal dichalcogenides [4-6, 10, 12], and 2D selenium [44]. The term "spectroscopic upconversion" is used to distinguish this effect from the well-known "photon upconversion", in which the sequential absorption of two or more photons results in the emission of light with a shorter wavelength than the excitation wavelength [45]. Thus, the observation of spectroscopic upconversion in UV-Vis TA spectroscopy clearly indicates multiphoton pumping of carriers even if the low pump fluences are used [26, 28]. Note here that although the term "low pump fluence" is widely used, it introduces confusion into nonlinear optics, since all nonlinear optical phenomena are uniquely determined by the light intensity (power density), and not by the light fluence (energy density). Therefore, a "low pump fluence" such as, for example, 100 μJ cm$^{-2}$ for 100 fs pulses corresponds to a pulse peak intensity of 1 GW cm$^{-2}$, which is more than sufficient to account for multiphoton excitation. Moreover, the shorter the femtosecond pulse, the higher the peak intensity, so it is quite surprising that multiphoton pumping in femtosecond UV-Vis TA spectroscopy of semiconductors was only recently discovered [26, 37].

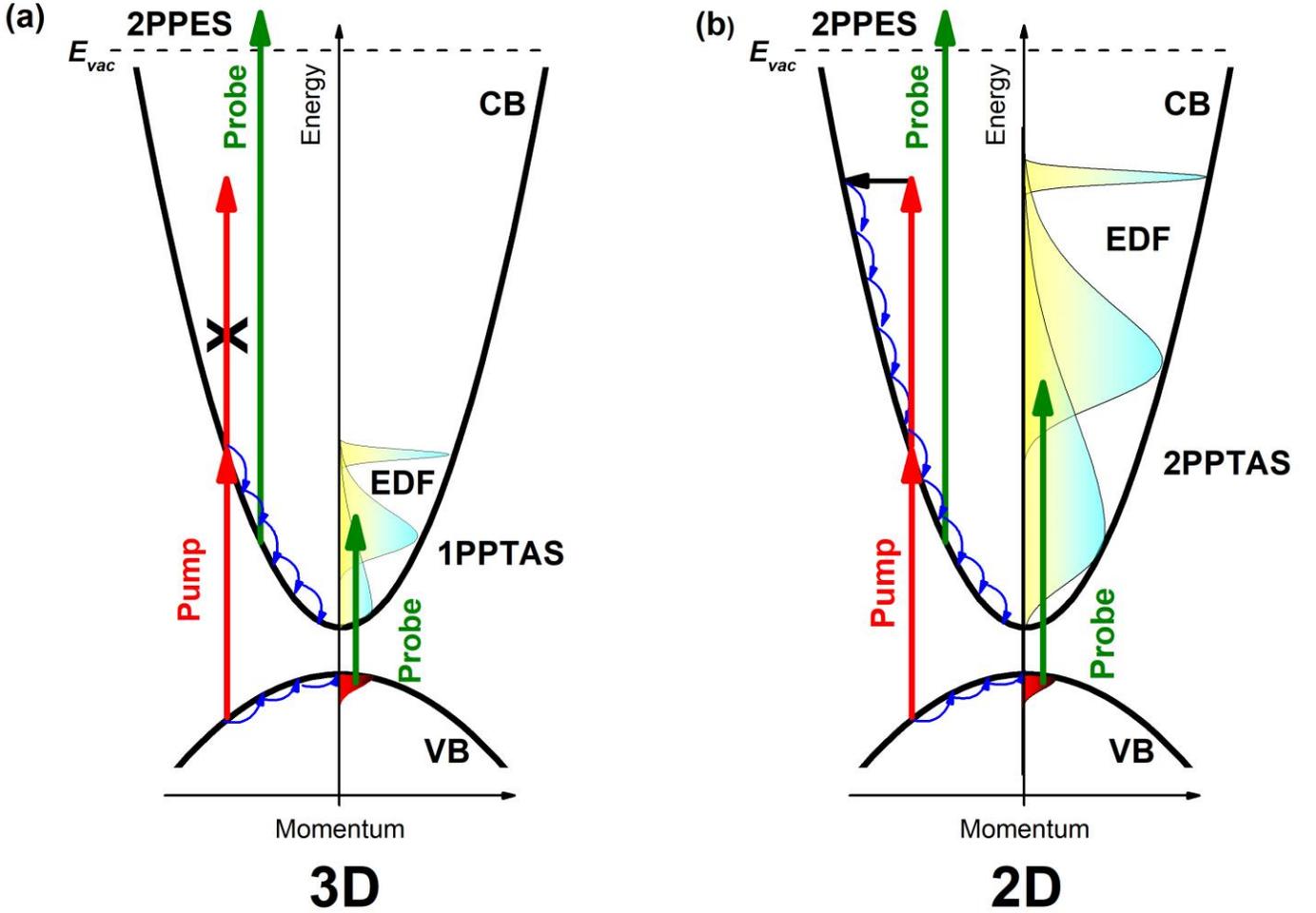

FIG. 1. Schematic band diagram of a semiconductor and comparison of two-photon photoemission spectroscopy (2PPES) with one-photon-pumped and two-photon-pumped transient absorption spectroscopy (1PPTAS and 2PPTAS, respectively) for bulk (3D) and 2D materials [(a) and (b) respectively]. The corresponding pump and probe transitions are shown with red and green arrows, respectively. The restriction of two-photon pumping (intraband absorption of the second photon) in 3D materials is shown in (a). This restriction is removed in 2D materials due to collisions of carriers with interfacial potential barriers, a process during which simultaneous conservation of energy and momentum is guaranteed [it is shown in (b) by the horizontal black arrow]. The cascade emission of phonons by hot electrons and holes in the conduction and valence bands (CB and VB, respectively) is shown schematically, along with the typical evolution of the electron energy distribution function (EDF).

However, since the pump photon energy exceeds the material's bandgap or is in resonance with excitonic states, such a multiphoton process requires electrons (holes) in the conduction (valence) band or excitons to absorb the pump light upon excitation. Despite the apparent simplicity of this statement, such an intraband process for bulk semiconductors is not allowed in the near-IR and UV-Vis ranges due to restrictions on the simultaneous conservation of energy and momentum [Fig. 1(a)] [46]. Consequently, the mechanism of multiphoton pumping in 2D materials has been proposed as involving the absorption of light by free carriers via inverse bremsstrahlung [26-29, 37, 47]. Such a process requires consideration of collisions between free carriers and the interfacial potential barriers of 2D materials [48–50] [Fig. 1(b)], a process that seems so natural in materials of this type. This mechanism typically occurs below the femtosecond optical breakdown threshold in condensed matter, gases, and liquids [51, 52] and removes any restrictions on energy-momentum conservation for intraband optical transitions.

This review presents experimental results on UV-Vis TA spectroscopy of various 2D materials for which TA spectra extend to energies significantly higher than the pump photon energy. We emphasize that this type of spectroscopic upconversion is associated with the dynamics of electrons initially excited by a multiphoton (multistep) process into high-energy states of the conduction band. The corresponding Pauli blocking within the electron relaxation dynamics provides the transient absorption bleaching, clearly manifested in the UV-Vis TA spectra [26-29, 37, 51-54]. The second contribution manifested in UV-Vis TA spectroscopy of 2D materials is bandgap renormalization. This process mainly affects the band edge states, but its high-energy tail also overlaps with and significantly modifies the absorption bleaching peaks [37, 53-56]. We also consider all other effects that can appear in multiphoton-pumped UV-Vis TA spectroscopy of 2D materials, such as inverse-bremsstrahlung-type free carrier absorption (FCA) [26, 28], valence band spin splitting due to the strong spin-orbit coupling and lack of inversion symmetry [4-8, 10-13, 57], and dynamic Rashba spin splitting of the conduction band induced by the built-in electric field [37].

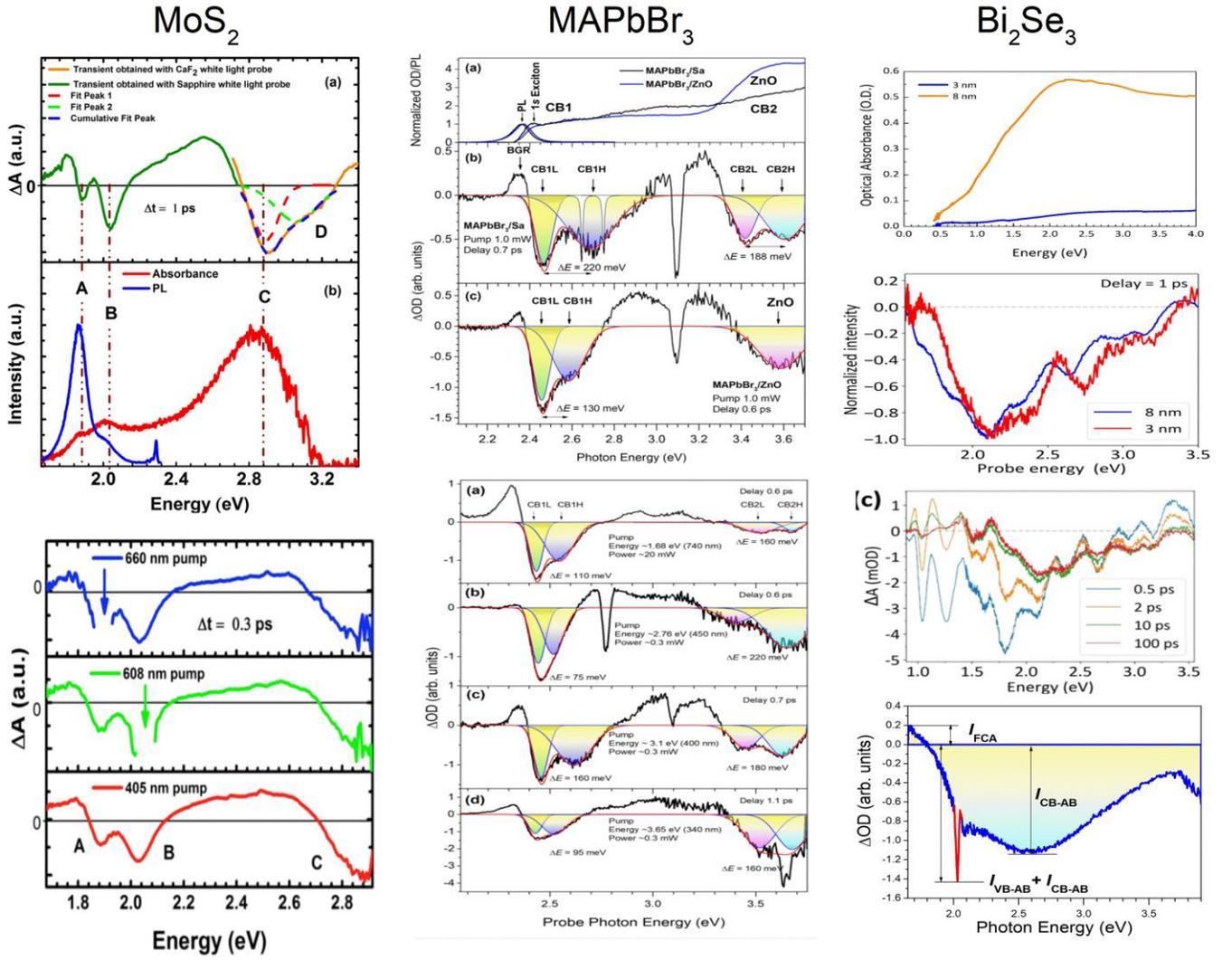

FIG. 2. Conventional absorption and transient absorption (TA) spectra of various 2D materials, as indicated at the top: left column – a monolayer of transition-metal dichalcogenide $MoS_2$, middle column – single-layer 20-nm-sized $MAPbBr_3$ nanocrystals, right column - topological insulator $Bi_2Se_3$ with a thickness of 3 and 8 nm (first, second and third figures from the top) and 10 nm (bottom figure). The left and middle columns also show the corresponding photoluminescence (PL) spectra as indicated, which were excited by photons with energies of 2.33 and 3.65 eV for $MoS_2$ and $MAPbBr_3$, respectively. For each TA spectrum shown in the left and middle columns, the pump photon energy at which it was measured is indicated. The TA spectra shown in the right column (second, third, and fourth from the top) were measured at pump photon energies of 2.48 eV, 0.62 eV, and 2.07 eV, respectively. All TA spectra were measured at room temperature, except for the third one from the top in the right column, measured at 77 K. The figures in the left column are adapted from [4] [Copyright (2016) American Physical Society]. The figures in the middle column are adapted from [37]. The figures in the right column are adapted from [24], except for the bottom one, adapted from [47]. All figures in the middle and right columns are taken from publications licensed under CC-BY 4.0, except for the bottom one in the right column [Copyright (2023) IOP Publishing].

We also emphasize that multiphoton pumping of carriers in 2D materials is due to collisional heating of electron-hole plasma caused by the inverse-bremsstrahlung-type FCA [26-29, 37, 47-49]. We compare this spectroscopy with two-photon photoemission spectroscopy (2PPES) of semiconductors [14-23, 58, 59] and multiphoton photoemission spectroscopy of noble metal surfaces [60–62]. Accordingly, we recognize that due to the band offsets at the interfaces of 2D materials [37, 63-65], absorption of pump photons by free electrons (holes) in the conduction (valence) band occurs because of collisions between free carriers and interfacial potential barriers [50]. Since this process is maximized at the peak intensity of the pump pulse, it precedes all carrier relaxation processes, including carrier-carrier thermalization, Auger heating and Auger recombination, carrier multiplication, carrier-phonon scattering, and electron-hole recombination [26, 66-70]. Finally, we conclude that multiphoton-pumped UV-Vis TA spectroscopy most accurately accounts for the ultrafast dynamics of carriers in 2D materials, since it includes their collisions with interfaces.

## 2. UV-Vis TA spectra and band assignment

*2.1. Typical UV-Vis TA spectra of 2D materials*

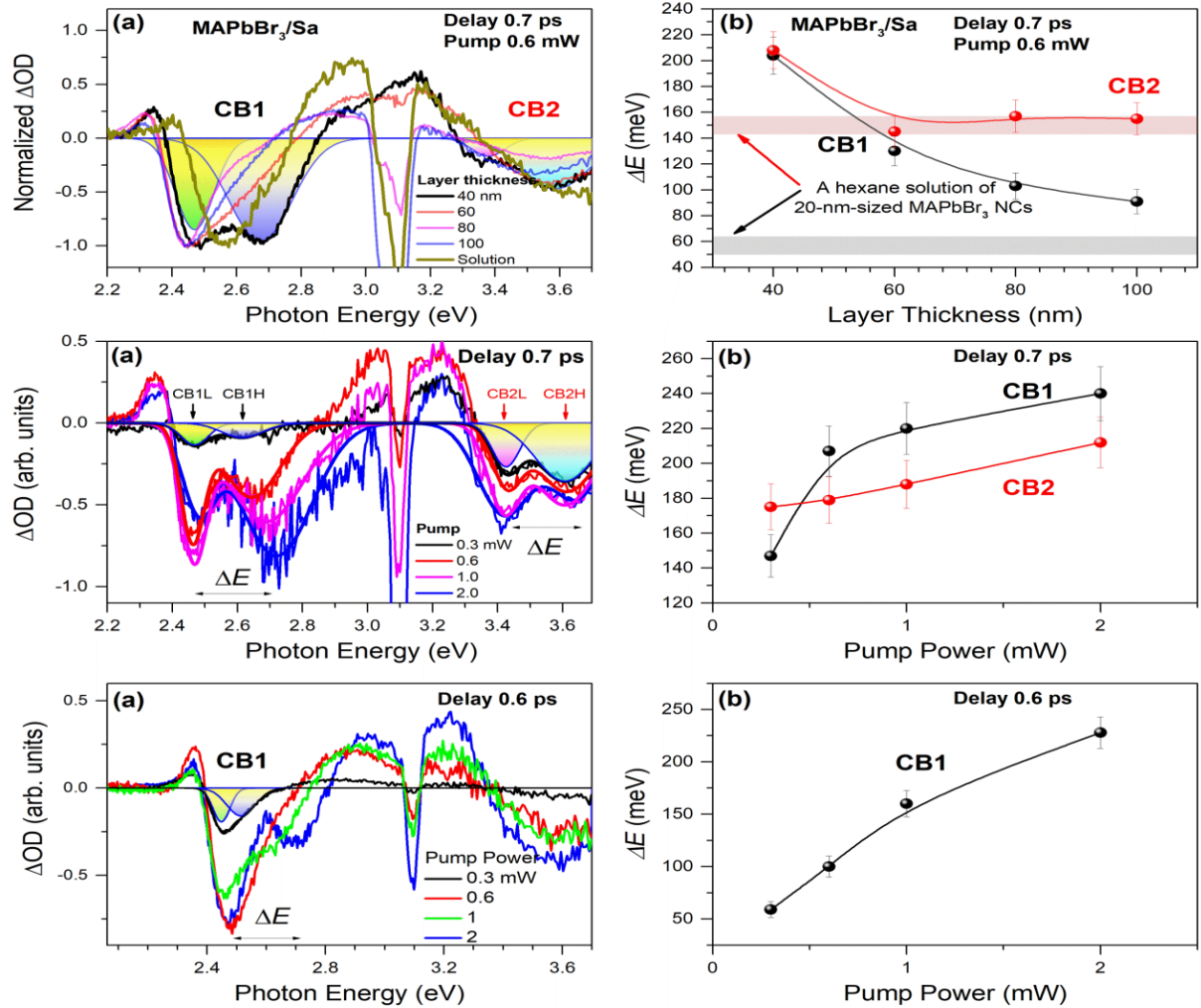

FIG. 3. Left column – The upper figure shows UV-Vis TA spectra of a hexane solution of MAPbBr$_3$ nanocrystals with a size of ~20 nm and the corresponding layer of MAPbBr$_3$ nanocrystals on a sapphire substrate with different layer thicknesses, as indicated by the corresponding colors. TA spectra were measured at ~0.7 ps delay using ~3.1 eV pumping with an average pump power of ~0.6 mW. The figures in the second and third rows show the TA spectra of MAPbBr$_3$ nanocrystals (layer thickness ~40 nm) on a sapphire and ZnO substrate, respectively, measured at the delay time as indicated, using ~3.1 eV pumping of various powers, as indicated by the corresponding colors. Right column - the corresponding Rashba spin splitting energy (ΔE) as a function of layer thickness and average pump power for the two conduction bands (CB1 and CB2), as indicated. All figures are adapted from [37], published under CC-BY 4.0 license.

Figure 2 compares the conventional and transient absorption spectra of several 2D materials that exhibit different bandgap energies. If the bandgap energy lies in the visible range, as in a monolayer of MoS$_2$ and single-layer MAPbBr$_3$ nanocrystals, similar negative and positive features are observed in all UV-Vis TA spectra, despite slightly different energy ranges. The appearance of these features is practically independent of the energy of the pump photon, regardless of whether macroscopic [4, 6, 10, 12, 37] or microscopic [5] experimental arrangements were used. In general, the negative features are associated with absorption bleaching (a pump-induced decrease in absorption) [4-6, 10, 12, 24, 26-29, 37, 53-56]. Because the absorption bleaching contributions exactly match all the characteristics of conventional absorption spectra, they mainly correspond to the density of states of the conduction band of 2D materials and substrates (e.g., ZnO [37]) [26]. This coincidence of spectral features in the conventional and transient absorption spectra clearly confirms that the negative peaks in the UV-Vis TA spectra are associated with the carrier population dynamics (filling of the conduction band states) and the corresponding Pauli blocking. Despite the general similarity, the splitting of negative peaks has a different nature. In a MoS$_2$ monolayer, the splitting is caused by two types of excitons (usually denoted A and B), arising due to the lack of inversion symmetry in a quantum system with strong spin-orbit coupling [57]. The resulting splitting (~160 meV) occurs at the maximum of the valence band [4-6, 10, 12]. The higher energy peak also belongs to the exciton (usually denoted C), although it is separated from the exciton peaks A and B by almost 1 eV and is much broader than them, probably due to the presence of an additional component (D) (Fig. 2, left column) [4].

On the contrary, in single-layer MAPbBr$_3$ nanocrystals, the splitting, although of a similar magnitude, is associated with dynamic Rashba splitting induced by the built-in electric field in two conduction bands (CB1 and CB2) [37]. The development of the built-in electric field in 2D materials is initiated by the photo-Dember field arising from the instantaneous charge separation between photoexcited carriers. The latter assignment is clearly confirmed by the fact that the splitting magnitude strongly depends on the nature of the substrate, gradually increases with increasing pump power, and begins to appear and gradually increase as the thickness of the MAPbBr$_3$ nanocrystal layer approaches the single-layer limit (Fig. 3) [37].

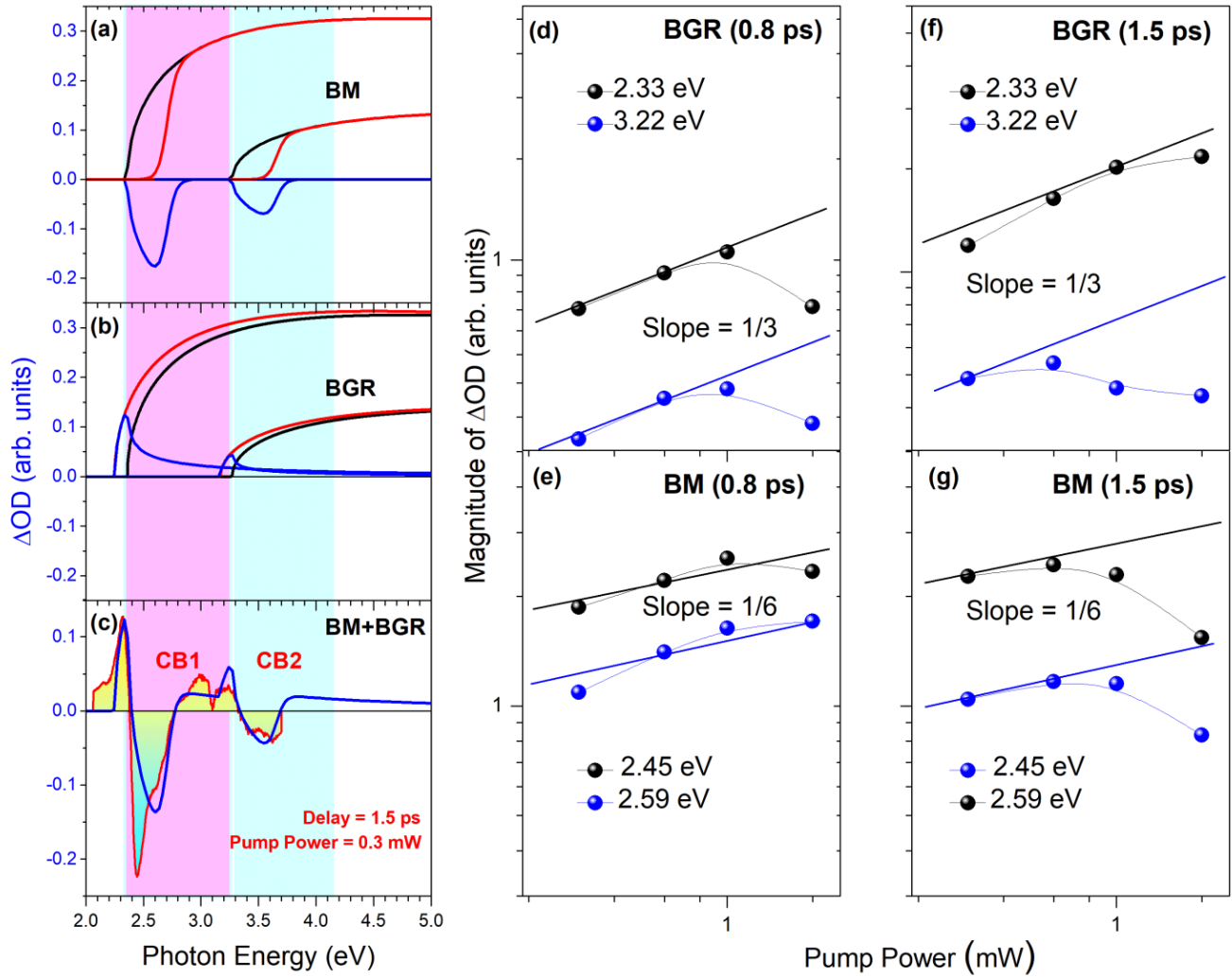

FIG. 4. [(a)–(c)] The numerical modeling of the Burstein-Moss (BM) and bandgap renormalization (BGR) contributions in UV-Vis TA spectra for two conduction bands (CB1 and CB2) of single-layer 20-nm-sized MAPbBr$_3$ nanocrystals. The black and red curves in (a) and (b) show the initial conduction band edge and the edge modified by the photoexcited carrier population, respectively (Rashba spin-splitting is ignored here), whereas the blue curves represent the difference between those two. The total effect (BM + BGR) is compared in (c) to the TA spectrum of single-layer MAPbBr$_3$ nanocrystals measured using ~3.1 eV pumping and parameters, as indicated. [(d)–(g)] The log–log plots of the average pump power dependences of the TA peak magnitudes (indicated by the corresponding colors), which were assigned to BM and BGR effects and measured at delay-times indicated in the corresponding brackets. The theoretically predicted slopes are shown as the straight solid lines. The figure is adopted from [37] published under CC-BY 4.0 license.

The weaker positive contribution in UV-Vis TA spectra characterizes the pump-induced increase in absorption and is due to renormalization of the bandgap of 2D materials [10, 37, 53–56]. Although this effect is most pronounced in the edge states, it is characterized by a high-energy tail that overlaps with the absorption bleaching peaks, thereby providing positive contributions energetically located below and above them (Figure 4). The appearance of a broadband positive contribution to the UV-Vis TA spectra of 2D materials means that, in addition to renormalization of the bandgap, photoexcited carriers have an integral effect on the entire conduction band [37, 54].

If the bandgap energy of 2D materials lies in the IR range, for example, in TI Bi$_2$Se$_3$, then the broadband negative contribution to the UV-Vis TA spectra is also associated with absorption bleaching. Because this contribution in most cases follows the general smooth trend of conventional absorption spectra (Fig. 2, right column), it also corresponds to the density of states of the conduction band and reflects the carrier population dynamics [26, 28, 47].

An additional positive feature of the UV-Vis TA spectra of 2D materials with bandgap energy in the IR range is related to the inverse-bremsstrahlung-type FCA (Fig. 2, right column). However, this effect is rarely observed, since the unique channel of carrier relaxation and a high spectral intensity of the supercontinuum probe beam are required, as was shown for 2D TI Bi$_2$Se$_3$ [26–29, 47].

## 2.2. Band assignment in the UV-Vis TA spectra

Thus, we have briefly discussed all the possible effects that can appear in the UV-Vis TA spectra of 2D materials and highlighted three unique contributions: absorption bleaching, bandgap renormalization, and inverse-bremsstrahlung-type FCA. We note here that these band assignments in the UV-Vis TA spectra of 2D materials are fundamentally different from those usually used for UV-Vis TA spectroscopy of chemical compounds and gas molecules. In the latter case, ground state bleaching, stimulated emission, excited state absorption, and product absorption are usually considered [71]. Correspondingly, in the Franck–Condon approximation, the intensity of the vibronic transition between the ground and excited states (or between the excited and other excited states with higher energy) is proportional to the square of the overlap integral between the vibrational wave functions of the two states involved in the optical transition. Obviously, this approach and the terminology associated with localized molecular orbitals are not applicable to 2D semiconductors. The reason is that delocalized free carriers populate nearly continuous energy levels of the conduction/valence band according to the Fermi-Dirac distribution, including a variety of many-body effects such as correlated motion of carriers and their interactions with ionized impurities and lattice phonons [46, 50]. Therefore, UV-Vis TA spectroscopy of 2D materials should be considered within the framework of condensed matter physics. Despite the apparent obviousness of this statement, there are many publications in which the UV-Vis TA spectra of 2D materials are interpreted using models and terminology typically applied to chemical compounds and gas molecules. To be on the right track and avoid terminological confusion, here we highlight in more detail the general trends in ultrafast carrier dynamics in 2D materials, including three main effects that could potentially contribute to their UV–Vis TA spectra.

### 2.2.1. Absorption bleaching

The absorption bleaching process [also known as Burstein-Moss (BM) shift or Pauli blocking] dynamically expands the bandgap of a semiconductor by filling the conduction/valence band with photoexcited carriers [37, 53-56]. As mentioned in the previous section, this process is responsible for the negative contribution to the UV-Vis TA spectra. The initial population of carriers excited by a femtosecond pump pulse exists as an electron-hole plasma. Because plasma is electrically neutral, it does not lead to any change in the complex refractive index of the materials and hence in their optical properties. Once the electron-hole plasma is thermalized due to carrier-carrier interactions, the carriers acquire their electron (hole) temperature causing the corresponding Fermi-Dirac distribution. The thermalization process leads to a corresponding separation of charges, and the resulting electric field changes the complex refractive index. The latter dynamics manifest themselves as an increase in the pump-probe signal. Further decay of the signal reflects the cooling of pump-excited carriers, which occurs due to their inelastic scattering by LO phonons (Fröhlich relaxation mechanism for polar 2D semiconductors [26, 32, 37, 72-74] and non-polar optical phonon scattering for non-polar 2D semiconductors [75]). The cooling dynamics are quite fast (<2 - 3 ps) and, in general, do not depend on the density of photoexcited carriers ($n_e$) [32, 37, 53-56]. Due to Pauli blocking, these relaxation dynamics, as well as the resulting accumulation of carriers at the edge of the conduction/valence band are imaged by the supercontinuum probe beam [4-6, 10, 12, 24, 26, 28, 37, 47]. In general, this behavior characterizes the electron EDF and manifests itself in a temporary redistribution of intensity within the absorption bleaching band of UV-Vis TA spectra towards lower energies [26, 37]. Consequently, the optical bandgap for the probe light appears to be dynamically expanded since the edge states become fully occupied by the pump-excited carriers (Fig. 4) [37, 53-56]. The peak amplitude of the absorption bleaching band weakly depends on the density of photoexcited carriers and scales as $n_e^{1/6}$ (Figs. 4, 5) [26, 28, 37].

### 2.2.2 Bandgap renormalization

In addition to population dynamics, there is an integral influence of pump-excited carriers on the band structure of 2D materials, which manifests itself in a broadband positive contribution to the UV-Vis TA spectra. The broadband trend maximizing slightly below the bandgap energy indicates that new unoccupied states are generated by a shift of the entire conduction band (i.e., a rigid shift of the conduction band without changing the effective electron mass) (Fig. 4) [37, 54]. This behavior is due to many-body effects and leads to bandgap narrowing, which was originally associated with the bandgap renormalization (also known as bandgap shrinkage) [37, 53-56]. Figure 4(b) shows the effect for the trivial case of the density of states in the form of $(E - E_0)^{1/2}$ for the parabolic band [46], although the real shape of the high-energy tail of the positive contribution reflects all the features of the conduction band, including its nonparabolicity [37, 54]. The peak amplitude of the bandgap renormalization effect scales with the photoexcited carrier density as $n_e^{1/3}$ (Figs. 4 and 5) [26, 37, 54].

Thus, the mentioned effects of absorption bleaching, and bandgap renormalization completely control the optical bandgap of pumped 2D materials for probe light and therefore clearly appear in their UV-Vis TA spectra [4-6, 10, 12, 26-29, 37, 47, 53-56].

### 2.2.3 Free carrier absorption

The most rarely observed contribution to the UV-Vis TA spectra of 2D materials is the absorption of light by free carriers. This effect provides another positive contribution to the UV-Vis TA spectra, in addition to bandgap renormalization. The reason this behavior is so rarely observed is that free (non-interacting) electrons (holes) in the conduction (valence) band cannot be excited to higher energy band states by absorbing light (an intraband optical transition) due to restrictions on simultaneous conservation of energy and momentum [46]. However, there are two exceptions that allow absorbing light by free carriers. The first is associated with their collective excitation (spectral region below the plasma frequency - Drude-type FCA) [46]. The second exception is the absorption of light by free carriers when they collide with impurities, lattice defects or with interfacial potential barriers [50] (the spectral region above the plasma frequency - the inverse-bremsstrahlung-type FCA) [26-29, 37, 47-49]. Drude-type FCA for the UV-Vis spectral range can be neglected, since the plasma frequency even at high densities of photoexcited carriers in semiconductors (~$10^{20}$ cm$^{-3}$) lies in the IR spectral range [37, 46].

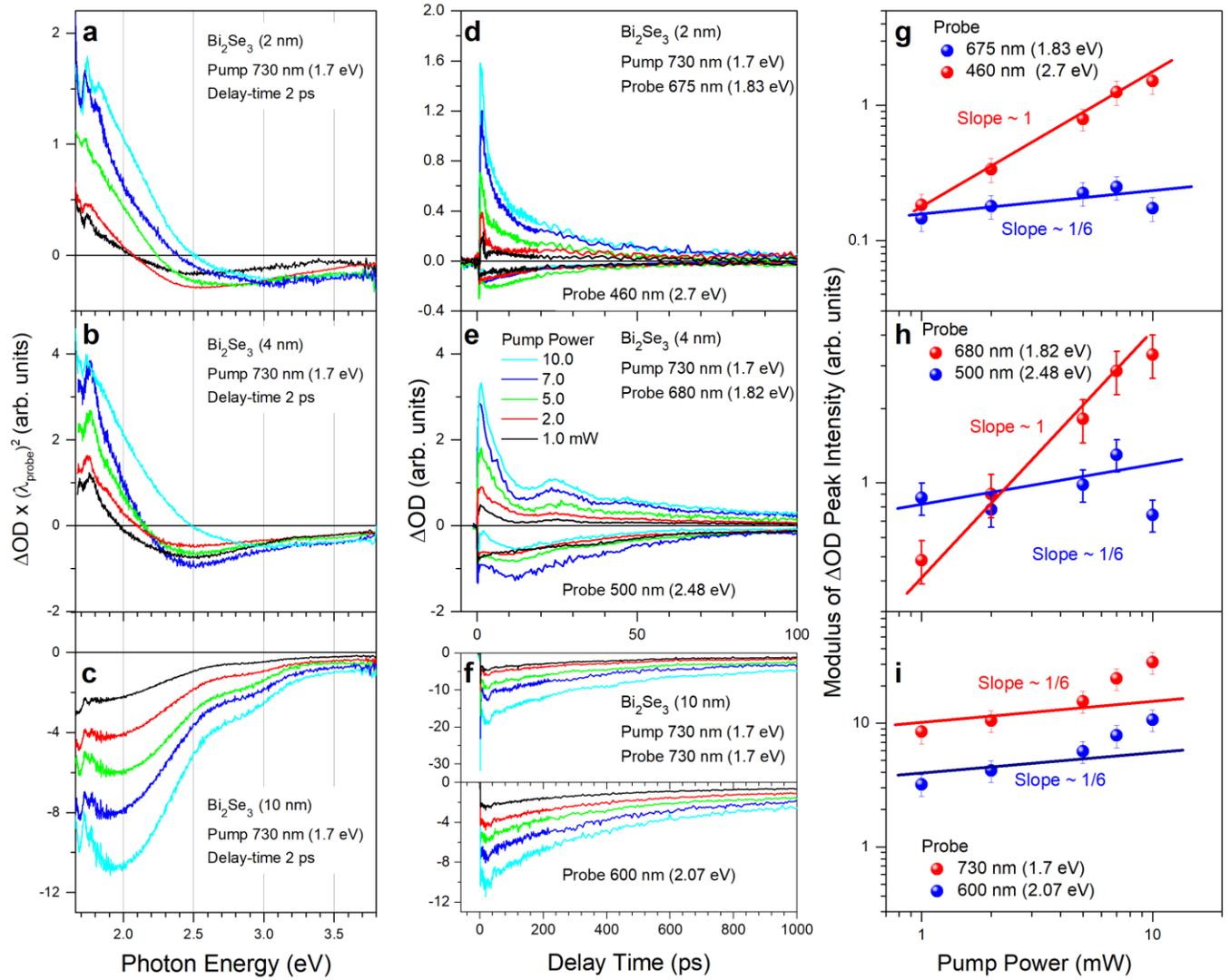

FIG. 5. (a−c) Set of UV-Vis TA spectra for Bi$_2$Se$_3$ films with various thicknesses, as indicated, measured at different average pump powers, as indicated by the corresponding colors. (d−f) Set of the corresponding pump-probe traces. All TA spectra and traces were measured under conditions indicated for each of the panels. (g), (h), (i) Power dependences of the modulus of the peak intensity of the pump-probe traces shown in (d), (e), and (f), respectively, which were plotted using log-log scales. The theoretically predicted slopes are shown by the corresponding color straight lines. The figure is adopted from [26] [Copyright (2021) American Chemical Society].

To observe the inverse-bremsstrahlung-type FCA, both a high density of photoexcited carriers and an intense probe laser beam (~$10^{11}$ W cm$^{-2}$) are required [51, 52]. Consequently, the reason that this process is rarely observed in UV-Vis TA spectroscopy of 2D semiconductors is that the spectral intensity of the supercontinuum probe beam is usually too low for the inverse bremsstrahlung process. Nevertheless, it was observed in the UV-Vis TA spectra of the topologically trivial phase of 2D TI Bi$_2$Se$_3$ (film thickness below 6 nm) (Fig. 5) [26, 28]. The observed exceptional behavior in the Dirac surface states is due to the dynamic accumulation of Dirac fermions in the surface states and hence their direct interaction with the initial (unattenuated) intensity of the supercontinuum probe light. Note also that the peak amplitude of the positive contribution to the UV-Vis TA spectra due to the inverse-bremsstrahlung-type FCA is linearly dependent on the density of photoexcited carriers. As a result, this contribution can be easily distinguished from the positive contribution associated with bandgap renormalization (Figs. 4 and 5).

### 3. Optimal experimental configuration

Compared with traditional optical spectroscopy, the overall performance of UV-Vis TA spectroscopy of 2D materials varies greatly depending on the actual design of the experimental setup and the angle of incidence of the pump and supercontinuum probe beams. Consequently, experimental results obtained by different research groups usually differ significantly from each other, not only due to differences in the quality of the samples, but also due to the actual design of the optical setup. It is worth noting that all commercially available TA spectrometers are usually designed for studying solutions of chemical compounds or nano/microcrystalline colloids. Since solutions and colloids are typically placed in a cuvette several millimeters thick, to maximize the amplitude of the pump-probe signal, it is necessary to increase the spatial overlap between the two focused beams. As a result, the angle between the pump and probe beams is usually set to a few degrees. In addition, due to the isotropy of solutions, the angle of incidence of the pump and probe beams is usually chosen arbitrarily. Obviously, such an experimental setup requires significant improvement for the study of 2D materials, usually presented in the form of thin films a few nanometers thick.

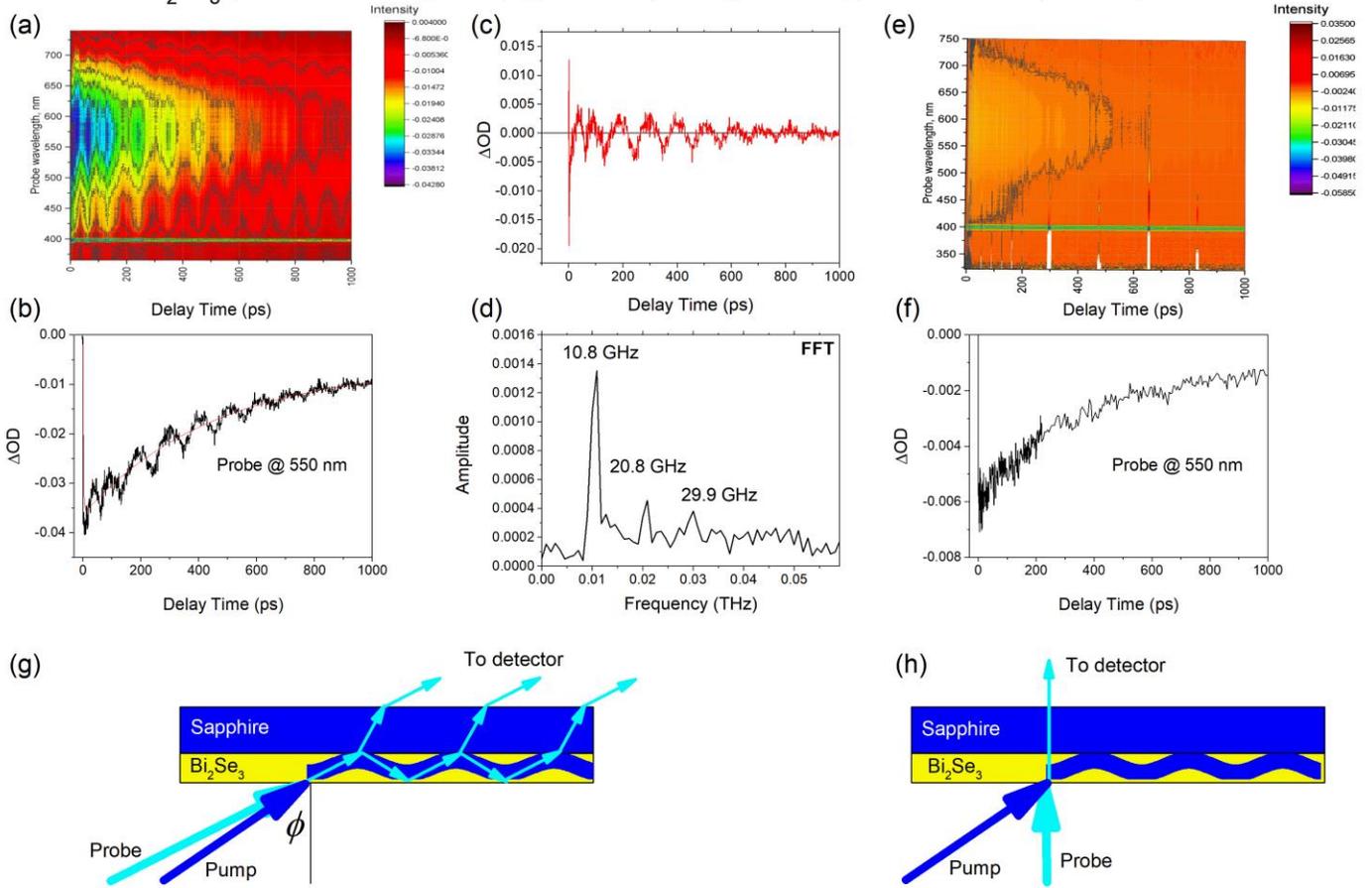

FIG. 6. (a) and (e) Representative pseudocolor TA spectral maps of a ∼10 nm thick $Bi_2Se_3$ film that were measured using ∼400 nm pumping (∼3.1 eV photon energy) with ∼2.0 mW average power in experimental configurations shown in (g) and (h), respectively. The color bar is shown to the right. (b) and (f) The corresponding pump-probe traces from (a) and (e), respectively, measured at a probe wavelength of 550 nm. (c) and (d) The oscillatory part extracted from the pump-probe trace shown in (b) and the corresponding fast Fourier transform (FFT), respectively. The frequencies of oscillations are marked in (d). For the experimental configurations shown in (g) and (h), the pump pulse excites a mixed longitudinal-transverse (shear) strains in the $Bi_2Se_3$ film, which can only be detected when using a non-normal incidence Brillouin configuration ($\phi \neq 0$).

Figure 6 shows an example illustrating this behavior for a 10 nm thick TI $Bi_2Se_3$ film. It is clearly seen that additional GHz range modulation of the UV-Vis TA spectral map can be achieved by changing the angle of incidence of the pump and probe beams. Specifically, oscillations occur exclusively at large angles of incidence, which is fully consistent with a non-normal incidence Brillouin scattering geometry [76, 77]. The dominant modulation frequency (~10.8 GHz) is approximately 4 times lower than the longitudinal acoustic wave cavity frequencies observed in thin $Bi_2Se_3$ films, whereas for a 10 nm thick film it should be at least 10 times higher [78]. However, this type of oscillation has been shown to be completely suppressed when normal incidence pumping is used for films less than 15 nm thick due to the Lamb wave excitation limit [78]. The suppression of oscillations with decreasing film thickness indicates a different regime achieved in acoustic wave dynamics due to the loss of flexural rigidity of the film caused by progressive shear lattice strains. The resulting intersurface coupling leads to joint out-of-phase/in-phase (symmetrical/asymmetrical relative to the mid-plane of the film) movements of both film surfaces, which is associated with the excitation of a Lamb wave. Thus, the observed low-frequency oscillations at high angles of incidence of the pump and probe beams indicate that the probe beam is modulated by fluctuations in dielectric constant, which ultimately arises due to strains created by a mixture of longitudinal and transverse sound waves. Obviously, when applying a normal incidence Brillouin scattering geometry, the oscillations become invisible (Fig. 6).

*3.1 Optimal experimental setup*

The optimal experimental setup for multiphoton-pumped UV-Vis TA spectroscopy of 2D materials appears to use the normal incidence of the supercontinuum probe beam to avoid any unwanted interference and scattering effects, as well as material-related distortions [26-29, 37, 47]. This probe beam orientation is commonly used in all commercially available UV-Vis and IR absorption spectrometers. In addition, the angle of incidence of the pump beam must be large enough (30 - 45 degrees) to reduce the spatial overlap between the pump and probe beams. This arrangement will also make it possible to minimize the intensity of artifacts emanating from the substrate and to simplify the procedure for suppressing scattered pump light in the detection channel of the setup. The spot size of the probing beam should be ~100 μm, but at the same time 2-3 times smaller than the spot size of the pump beam, to avoid a mismatch between the peak intensities of the pump and probe beams and to minimize the influence of material inhomogeneity. It should also be noted that the UV-Vis TA spectra obtained using transient reflection spectroscopy followed by Kramers-Kronig transformation are less accurate due to double-pass backscattering at the 2D-material-substrate interface. Therefore, preference is given to measuring the transmission of the probe light rather than its reflection.

We also note that everything that has been said about the macroscopic experimental arrangement of multiphoton-pumped UV-Vis TA spectroscopy is equally applicable to the microscopic one, despite the normal incidence of both the pump and probe beams. In addition, much lower laser powers must be used to avoid the femtosecond optical breakdown threshold of the materials when using laser spots of a few micrometers in size.

*3.2 Chirp correction*

One of the fundamental problems of UV-Vis TA spectroscopy of 2D materials is that the TA spectra measured in the subpicosecond range are usually convolved with the chirp of the supercontinuum probe beam. Since UV-Vis TA spectra at this timescale characterize ultrafast carrier relaxation dynamics associated with the LO-phonon scattering mechanism [26, 32, 37, 72-74], the chirp extraction procedure becomes extremely important to obtain accurate data. The chirp effect in commercially available TA spectrometers is usually minimized by using parabolic mirrors to collimate and focus the supercontinuum probe beam, as well as by optimizing the geometry of the experiment. However, the remaining chirp is still present, and its influence varies depending on the actual design of the experimental setup. This behavior means that the chirp of the supercontinuum probe beam must be measured and corrected numerically as part of a post-experimental procedure for each specific experimental setup. To measure the chirp effect, one can use degenerate four-wave mixing or two-photon absorption in the substrate measured at the same experimental conditions [28, 29, 79]. It is worth noting that incorrect chirp deconvolution often leads to the complete exclusion of the ultrafast relaxation stage associated with electron-phonon interaction from the UV-Vis TA spectra.

*3.3 UV-Vis TA spectrum sign*

In UV-Vis TA spectroscopy, probe light is pathing through the sample, and then collected and monitored using a spectrograph coupled with lock-in amplifier and CCD detector. Within this experimental configuration, the sign of the signal obtained by amplitude modulation of the pump beam can be easily switched to the opposite by changing the modulation phase of the lock-in amplifier by 180 degrees. Thus, the sign of TA spectra is ambiguous and must be chosen considering the nature of the physical process that causes the corresponding optical response. An incorrect choice of the sign of the UV-Vis TA spectrum will certainly lead to an incorrect interpretation of the experimental data.

## 4. The basic concepts of multiphoton-pumped UV-Vis TA spectroscopy

As we mentioned in Section 2.2.3, the inverse-bremsstrahlung-type FCA in UV-Vis TA spectra was observed only for 2D TI $Bi_2Se_3$, since these materials exhibit a unique carrier relaxation channel through Dirac surface states. The main reason that this kind of absorption is rarely observed is apparently due to the rather low spectral intensity of the supercontinuum probing beam. However, this FCA mechanism can be very efficient for a more intense narrowband pump beam. Since the amplitude of the inverse-bremsstrahlung-type FCA signal linearly depends on the density of photoexcited carriers and, therefore, on the pump power (Fig. 5) [26, 37, 49], the first-order optical process can be expected to dominate the pumping dynamics. In this regard, the inverse-bremsstrahlung-type FCA has been suggested to be the main phenomenon determining the implementation of multiphoton-pumped UV-Vis TA spectroscopy of 2D materials [26, 28, 37]. However, there are a few fundamental points that need to be addressed, as discussed below.

The pump regime in multiphoton-pumped UV-Vis TA spectroscopy of 2D materials largely corresponds to the regime in 2PPES of semiconductors [14–23, 58, 59]. However, photoexcited electrons remain in the 2D material or are trapped at its interfaces rather than leaving the sample [26–29, 47]. Although 2PPES is usually associated with a two-photon process, it is more of a two-step process that uses a one-photon process at each step. In particular, the first photon creates a nonequilibrium distribution of electrons in the conduction band of a semiconductor, which is probed by the second photon through photoemission. The first process is associated with the excitation of electrons from the valence band to the conduction band and is therefore determined by the linear (first order) susceptibility of the material. In contrast, the second photon interacts with the electron gas and allows the photoexcited electrons to reach high-energy states of the continuum above the vacuum level. The interaction of a light pulse with the electron gas is also characterized by linear susceptibility, but of a completely different nature and scale. It is worth noting that despite the two-step nature of the 2PPE process, it can sometimes be characterized as a one-step process using effective second-order susceptibility [58]. However, this approach is in sharp contrast to traditional nonlinear optics, where the two-photon absorption process is usually characterized by third-order susceptibility [80, 81]. Moreover, the one-step approach in 2PPES is not justified if the sum frequency of the two photons is close to the plasmon resonance [58].

The high-energy continuum in 2PPES means that an electron with any crystal momentum can find a corresponding high-energy state and leave the crystal, after which it is selectively analyzed for its initial energy and crystal momentum using angle-resolved photoemission. With this experimental approach in 2PPES, simultaneous conservation of energy and momentum for optical transitions is always observed [59]. By introducing a delay between two light pulses, a time resolution in 2PPES is provided that allows the dynamics of photoexcited electrons to be studied with high precision [14-23, 58, 59]. Specifically, the combination of angular and time resolution in 2PPES allows one to study the time evolution of EDF for a population of photoexcited electrons in the conduction band relative to their crystal momentum. Using this method, the dynamics of photoexcited electrons in the conduction band of 2D materials have been studied [14–23].

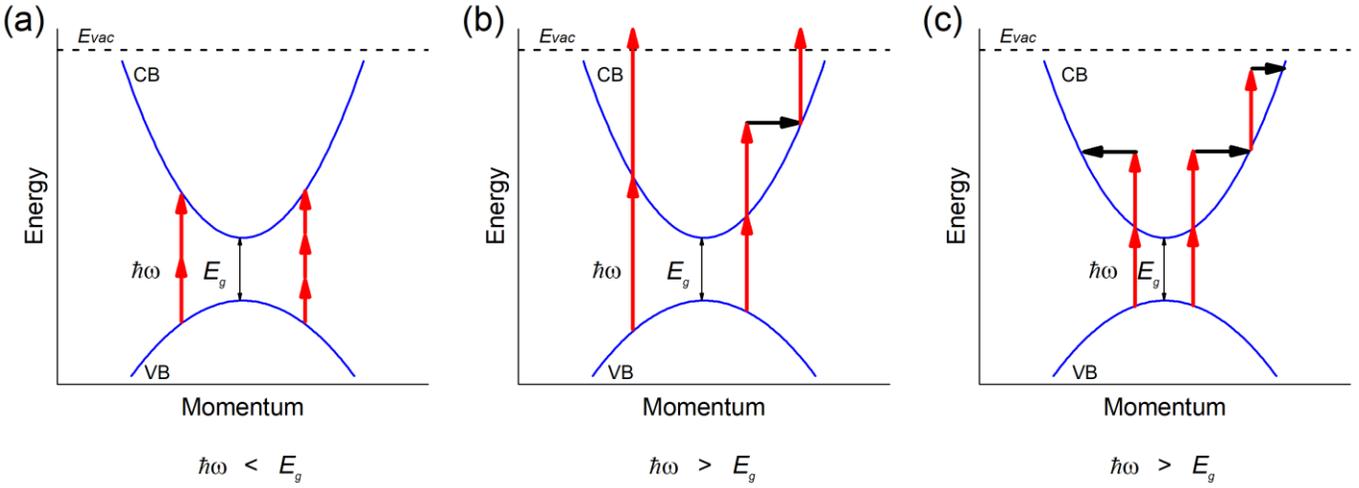

FIG. 7. (a) A schematic representation in momentum space of two-photon and three-photon absorption processes with photon energy ($\hbar\omega$) below the bandgap energy ($E_g$) of the semiconductor. (b) and (c) A similar to (a) representation of two-step and three-step photoemission and absorption processes, respectively, but with photon energies exceeding the semiconductor bandgap energy. $E_{vac}$ represents the energy of the vacuum level. Collisions of photoexcited carriers with potential barriers at the interfaces of 2D materials, indicated by horizontal arrows, are required to simultaneously conserve energy and momentum for intraband optical transitions.

Thus, to create an electron gas in the conduction band of a semiconductor, the photon energy of the first laser pulse in 2PPES must always exceed its bandgap. The situation is completely different when the energy of the incident photon is less than the bandgap energy of the material. In the latter case, the excitation process can be exclusively multiphoton, when to bridge the valence and conduction bands it is necessary to simultaneously absorb several photons [Fig. 7(a)] [80-85]. Multiphoton absorption processes are characterized by high-order susceptibilities [80, 81] and should therefore be distinguished from multi-step one-photon absorption processes, which are usually characterized by first-order susceptibilities at each step [37, 48, 49].

Unlike 2PPES, traditional pump-probe optical measurements non-selectively capture the influence of all possible elementary excitations in a semiconductor on its complex refractive index. This behavior leads to ambiguity in the interpretation of the results obtained. The situation can be significantly improved using UV-Vis TA spectroscopy [2-13, 24, 26-30, 37-40, 44, 47]. Despite the loss of electron momentum distribution, UV-Vis TA spectroscopy, like 2PPES, tracks population dynamics in the conduction band. Specifically, the time evolution of EDF for photoexcited electrons can be monitored directly during their relaxation but using the Pauli blocking mechanism [26-29, 37]. Such population dynamics are clearly visible in UV-Vis TA spectroscopy, especially for high-energy states of the conduction band, where other mechanisms, such as bandgap renormalization, play a minor role [37].

However, it has recently been shown that the UV-Vis TA spectra of some 2D materials extend to energies significantly higher than the pump photon energy [4–6, 10, 12, 24, 26, 28, 37, 44]. This kind of spectroscopic upconversion clearly indicates multiphoton absorption of pump photons [26, 28, 37]. Since the energy of the pump photons also exceeds the bandgap of the 2D material or is in resonance with the exciton states, such a multiphoton process requires electrons (holes) in the conduction (valence) band or excitons to absorb the pump photons upon excitation. However, due to restrictions on simultaneous conservation of energy and momentum for optical transitions, direct absorption of light by an electron in the conduction band is allowed only when the final excitation step takes the electron into a high-energy continuum, as in 2PPES [Fig. 7(b)]. Alternatively, the electron must scatter its momentum to move to a higher energy state of the conduction band (an intraband optical transition) (Fig. 7) [46]. For this reason, multiphoton-pumped UV-Vis TA spectroscopy of conventional bulk semiconductors has never been considered.

The situation appears to be significantly different when electrons are excited in 2D materials. Since the depth of light penetration into the bulk counterparts of 2D materials and the mean free path of photoexcited carriers significantly exceed their thickness [31-36], it can be expected that photoexcited carriers will experience collisions with interfacial potential barriers in the presence of a strong laser field. The height of the potential barriers in 2D materials is determined by the band offsets at their interfaces [63-65]. The collision of electrons with potential barriers is confirmed by the fact that the energy of the electron after the absorption of a pump photon can exceed the height of the potential barrier, allowing the electron to be trapped at the interface [47]. If the carrier density is high enough and the laser light has sufficient power, collisions of electrons with interfacial potential barriers remove any restrictions on energy-momentum conservation for intraband optical transitions [50]. As a result, photoexcited electrons can be transferred to higher energy conduction band states [Fig. 7(c)]. This excitation mechanism is known as the inverse-bremsstrahlung-type FCA [26-29, 37, 47-49] and typically occurs during impact and avalanche ionization below the femtosecond optical breakdown threshold in condensed matter, gases, and liquids [51, 52].

Moreover, the inverse-bremsstrahlung-type FCA mechanism, in addition to the absorption mechanism occurring through intermediate resonances, is also responsible for heating the electron gas in noble metals [60-62]. It is well known that if the energy of an electron upon absorption of a photon exceeds the height of the potential barrier at the metal-medium interface, then the electron with some probability can be emitted from the metal [50]. As we mentioned above, a similar type of electron dynamics has also been proposed for multiphoton-pumped UV-Vis TA spectroscopy of 2D materials, where electron trapping at interfaces is considered instead of photoemission [47]. Thus, inverse-bremsstrahlung-type FCA is the key process responsible for the nonequilibrium dynamics of multiphoton-pumped electron gas in electronic systems such as noble metals and topological insulators. Figure 7 compares in momentum space all the mentioned electronic transitions associated with two- and three-photon absorption, two- and three-step photoemission, and two- and three-step absorption. These trends suggest the vertical electronic transitions (Franck-Condon principle) and the inverse-bremsstrahlung-type FCA mechanism. It is worth noting that, despite the obvious multistep nature of the pumping process, we continue to call them multiphoton, like the two-photon (two-step) process in 2PPES.

Thus, collisional heating of photoexcited electron-hole plasma due to the inverse-bremsstrahlung-type FCA leads to multiphoton pumping of carriers in 2D materials to energies significantly exceeding the energy of pump photons. This process completely determines multiphoton-pumped UV-Vis TA spectroscopy of 2D materials. It is worth noting that the collisional heating model is ideally suited to the nature of 2D materials, since the frequency of collisions between free carriers and the interfacial potential barrier is expected to vary inversely with the thickness of 2D materials. To estimate the time between collisions, we use a 2D material thickness of ~1 nm and a maximum Fermi velocity of $10^6$ m s$^{-1}$ [86]. The resulting value is 1 fs, a time that is well suited for the inverse-bremsstrahlung-type FCA in condensed matter [52].

This estimate implies that the collisional heating of carriers in 2D materials occurs within the pump pulse peak intensity. This behavior means that multiphoton excitation precedes all possible relaxation processes, including thermalization of carriers to electron (hole) temperature upon reaching the Fermi-Dirac distribution, Auger heating and Auger recombination, carrier multiplication, electron-phonon scattering, and electron-hole recombination [26, 66-70]. A similar conclusion can be drawn regarding the observed dynamics of ultrafast relaxation of carriers through the split subbands. Both giant spin-orbit splitting in monolayers of transition-metal dichalcogenides [57] and dynamic Rashba splitting in single-layer MAPbBr$_3$ nanocrystals [37] affect the dynamics of carrier relaxation, but not the dynamics of their excitation. The lack of influence of the multiphoton pumping on all relaxation processes explains why UV-Vis TA spectroscopy with multiphoton pumping was not recognized in earlier experiments.

This behavior also means that no relaxation process can be used to reveal the photonicity of the pump regime. As we noted in Sections 2.2.1 and 2.2.2, the pump power dependence of the amplitude of the TA signals associated with absorption bleaching or bandgap renormalization also cannot be used to estimate the photonicity of multiphoton pumping. The reason is that the amplitude of these TA signals is not directly proportional to the density of photoexcited carriers [26, 28, 37, 52], as happens, for example, for photoluminescence (PL) signals [80–85]. Thus, the only evidence of multiphoton pumping in UV-Vis TA spectroscopy is the extension of the actual spectral range of probing towards energies exceeding the energy of pump photons. From this point of view, multiphoton-pumped UV-Vis TA spectroscopy is almost identical to one-photon-pumped TA spectroscopy, but with an extended range over which the ultrafast dynamics of photoexcited carriers can be studied. As an example, we can consider the dynamic Rashba splitting in the higher energy conduction band (CB2) of single-layer MAPbBr$_3$ nanocrystals and the splitting of the C-exciton peak in a MoS$_2$ monolayer (Fig. 2). Depending on the energy of the pump photons, both dynamics can be observed either in the energy region below the energy of the pump photons (one-photon pumping) or in the energy region exceeding the energy of the pump photons (multiphoton pumping) [4, 37].

## 5. Recent applications of multiphoton-pumped UV-Vis TA spectroscopy

We are now reviewing several recent experimental observations for 2D materials, which can be associated with multiphoton-pumped UV-Vis TA spectroscopy. Note that UV-Vis TA spectra extending to energies significantly higher than the energy of pump photons were first identified for single-layer MAPbBr$_3$ nanocrystals [37] and thin-film TI Bi$_2$Se$_3$ [26, 28]. The observed ultrafast dynamics of photoexcited electrons in these 2D materials is attributed to their multiphoton (two- or three-photon) pumping into high-energy states in the conduction band followed by electron-phonon cooling. Moreover, it was concluded that the relaxation dynamics of photoexcited carriers under multiphoton pumping are very similar to the dynamics induced under one-photon pumping of the same energy [26–29]. This conclusion was also confirmed in the more recent application of multiphoton-pumped (three- to five-photon) UV-Vis TA spectroscopy to Bi$_2$Se$_3$ thin films [24]. All these observations confirm the estimate made in the previous section that multiphoton pumping occurs within the peak intensity of the pump pulse and does not affect any relaxation processes. In contrast, multiphoton-pumped UV-Vis TA spectra of 2D selenium show approximately an order of magnitude slower dynamics compared to the one-photon-pumped spectra [44]. This behavior was explained by different dynamics of carrier relaxation when probing surface and near-surface regions.

Although similar spectroscopic upconversion has also been observed for monolayers of transition-metal dichalcogenides, their UV-Vis TA spectra in the energy range exceeding the pump photon energy have not been explained as induced by multiphoton (multistep) pumping, regardless of whether macroscopic [4, 6, 10, 12] or microscopic [5] experimental configurations were used. Specifically, for MoS$_2$ monolayers, three negative peaks in the UV-Vis TA spectra were assigned to two spin-split excitons at the band edges (A and B excitons) and a hot carrier exciton in the "band nesting" region (C exciton) [4, 5, 11-13]. Exciton features are strongly pronounced in MoS$_2$ monolayers due to reduced dielectric screening and hence the enhanced Coulomb interactions. Accordingly, exciton binding energy in a MoS$_2$ monolayer of ~300 meV [87-89] is significantly higher compared, for example, with that for single-layer MAPbBr$_3$ nanocrystals ~35 meV [37]. The excitation of higher energy excitonic states (excitons B and C) when pumped into the lowest energy exciton state (exciton A) is explained by a significant coupling between excitonic resonances in momentum space [4, 5] or due to a "upconversion/many-body" process [12].

Regarding exciton coupling, we note that the term "coupling between exciton resonances" is probably borrowed from coherent four-wave mixing spectroscopy of semiconductor quantum wells [90]. In these quantum systems, the splitting of exciton states is too small (a few meV), so they are excited simultaneously within the laser pulse width. This behavior contrasts sharply with excitonic resonances observed for monolayers of transition-metal dichalcogenides, where, due to the large splitting energy, excitonic resonances are excited incoherently and independently. Accordingly, some complex models have been proposed to explain this unusual, highly energetic coupling between excitonic resonances [10, 12].

Since the energy difference between excitons A and B, as well as A and C in $MoS_2$ monolayers is about 160 meV (1857 K) and 1 eV (11606 K), respectively, the existence of extremely hot (nonequilibrium) excitons can be assumed. However, the latter value is more than twice the predicted maximum parameter for valence band splitting due to spin-orbit interaction [57] and at least three times the exciton binding energy [87, 88]. Therefore, even if such high-energy coupling between excitonic resonances A and B can be achieved through giant spin-valley coupling [91-94], coupling between excitonic resonances A and C (or B and C) seems unlikely due to the extremely large energy mismatch. In other words, such coupling requires a dramatic modification of the band structure by photoexcited A-excitons, comparable to that caused by quantum confinement when the layered system reaches the monolayer limit [95]. Such a "giant" renormalization of the bandgap was not observed in the UV-Vis TA spectra of monolayers of $MoS_2$ and other transition-metal dichalcogenides.

Note also that the 'upconversion/many-body' model makes sense for quantum dots (0D semiconductors) in the regime of strong spatial confinement [66-70]. Therefore, its application to 2D materials remains questionable since there is purely vertical quantum confinement. It has been shown that even for 1D quantum confinement (nanorods), the efficiency of many-body effects (such as carrier multiplication and Auger heating) drops significantly and strongly depends on the actual carrier density in 1D semiconductors compared to that in the 0D ones [66, 68]. Thus, no reasonable physical model has been proposed to explain the excitation of C-exciton states in $MoS_2$ monolayers when pumped into A-exciton states.

In addition, we draw attention to the fundamental difference between the UV-Vis TA spectra of a monolayer of $MoS_2$ and single-layer $MAPbBr_3$ nanocrystals, despite their overall similarity (Fig. 2). In particular, the decay traces of the three absorption bleaching peaks A, B and C in a $MoS_2$ monolayer show similar decay dynamics [4], suggesting that all types of excitons relax simultaneously at almost the same rates. Such a general relaxation of the excitons of different energies seems surprising, since there is a huge energy gap between them. In contrast, the decay traces of the higher energy split component in the UV-Vis TA spectra of single-layer $MAPbBr_3$ nanocrystals show a much shorter decay compared to the decay of the lowest energy component [37]. The latter behavior clearly demonstrates the general tendency of photoexcited carriers to relax into the lowest energy states at the band edges where they recombine. As a result, the onset of the corresponding pump-probe traces (the so-called zero time) is shifted towards negative times within the relaxation limits [29]. However, such traditional relaxation dynamics do not appear in the UV-Vis TA spectra of a $MoS_2$ monolayer [4], although according to theoretically predicted carrier relaxation rates [72], they should appear on time intervals of several hundred femtoseconds.

Moreover, it is sometimes possible to observe a slower cooling of hot C-excitons compared to the band edge excitons, which is attributed to the "favorable band alignment" and "transient excited-state Coulomb environment" [12]. We note again that the latter approach likely considers many-body interaction trends developed for quantum dots [66–70], and therefore some additional evidence is needed to account for such many-body effects in 2D materials. In addition, the slowdown of relaxation dynamics is usually associated with the low probability of multiphonon processes leading to the formation of a phonon bottleneck [72, 80]. This effect was underestimated in [12], although the interaction with lattice phonons (Fröhlich relaxation mechanism) dominates [72-74]. As in [4], the decay traces of A, B and C excitons presented in [12] do not show any zero-time shift. This behavior indicates the absence of relaxation of C excitons towards excitonic states with the lowest energy.

Moreover, the entire stage of ultrafast relaxation associated with electron-phonon scattering may be lost. This behavior is illustrated by the population of a higher energy exciton resonance (B) when pumped into a lower energy exciton resonance (A) and vice versa at zero delay [10]. Obviously, this observation completely ignores any carrier population dynamics in the subpicosecond range. We also note that even in coherent spectroscopy of four-wave mixing of semiconductor quantum wells, there is no signal at zero delay, as nevertheless presented in [10], and a certain delay between incident light pulses is necessary to populate excitonic resonances and observe the coherent coupling between them [90].

Despite this fundamental difference between the UV-Vis TA spectra of a $MoS_2$ monolayer and single-layer $MAPbBr_3$ nanocrystals, the discussed general trend of carrier relaxation is confirmed by the corresponding PL spectra. For both the $MoS_2$ monolayer and single-layer $MAPbBr_3$ nanocrystals, the PL is dominated by recombination of the lowest energy excitons, despite their excitation through the higher energy states at 2.33 and 3.65 eV, respectively (Fig. 2). This behavior clearly indicates relaxation of photoexcited carriers towards the lowest energy exciton states, including efficient spin-valley coupling between A and B excitons and the corresponding phonon-assisted intervalley spin relaxation [91-94].

Thus, the unusual relaxation dynamics observed in the UV-Vis TA spectra of $MoS_2$ monolayers contradicts the general trends of carrier relaxation in semiconductors. The similar relaxation rates observed for A, B, and C excitons in $MoS_2$ monolayers, the lack of relaxation to lower energy excitonic states, and the lack of subpicosecond relaxation dynamics are most likely due to inappropriate experimental conditions or inaccurate post-experimental processing of UV-Vis TA spectra and chirp correction, as discussed in Section 3.

An alternative interpretation of the UV-Vis TA spectra of monolayers of transition-metal dichalcogenides can be proposed based on multiphoton pumping, as discussed above for other 2D materials [24, 26, 28, 37, 44]. Accordingly, all contradictions in explaining the observed high-energy coupling between excitonic resonances can be eliminated. Following general trends, multiphoton-excited electrons and holes relax downward or instantaneously form excitons and then relax into band edge states depending on the exciton binding energy and the strength of the electron-phonon interaction. This Fröhlich relaxation mechanism of highly nonequilibrium carriers is also valid for monolayers of transition-metal dichalcogenides [72-74] and is generally accepted for polar III–V and II–VI quantum structures [96]. This model can explain all the above-discussed features of the dynamics of carrier relaxation in 2D materials, including monolayers of transition-metal dichalcogenides. Accordingly, through relaxation, multiphoton-pumped carriers populate all possible lower energy states in the conduction/valence band and exciton states (Fig. 1), as typically occurs in semiconductor quantum wells [96]. This mechanism does not require any coupling between such energetically distant exciton resonances in monolayers of transition-metal dichalcogenides. From this point of view, there is no difference between 2D materials and semiconductor quantum wells, except for the greater thickness and not so high potential barriers at the interfaces in the latter case.

Regarding the multiphoton-pumped UV-Vis TA spectra of TI $Bi_2Se_3$ thin films, we note that, despite the overall smooth trend of conventional and transient absorption spectra, some internal structure of the broad absorption bleaching band can be observed (Fig. 1, right column). The internal structure is mainly observed in UV-Vis TA spectra measured at low temperatures, although it also appears slightly at room temperature [24]. In view of the above-mentioned trends in the splitting of the valence and conduction bands and the continuous density of states in them, such an internal structure of the absorption bleaching band seems unrealistic. Since the band structure and inhomogeneity of the sample do not change with temperature, the observed internal structure indicates the specifics of the experimental setup rather than a general trend with decreasing temperature. As discussed above in Section 3, UV-Vis TA spectroscopy is strongly influenced by the experimental conditions. Therefore, even the cryostat windows, their quality and thickness, and the angle of incidence of the laser beams on them can affect the final performance. In addition, proper post-experimental processing of UV-Vis TA spectra is critical to obtain accurate data.

The interpretation of the results of multiphoton-pumped (two- and three-photon) UV-Vis TA spectroscopy of 2D selenium also looks doubtful [44]. Specifically, the positive broadband contribution to the spectra was explained by absorption in the excited state. It is worth noting that such a positive broadband contribution has never been observed in UV-Vis TA spectra for any other 2D materials. Moreover, the approach and terminology used in [44] are usually applied to chemical compounds and gas molecules (Section 2.2). Since the absorption of light by electrons in the conduction band is allowed only as the inverse bremsstrahlung process (Section 4), the proposed interpretation of the UV-Vis TA spectra for 2D selenium seems questionable. On the other hand, if we assume that the sign of the contribution was incorrectly determined, as discussed in Section 3.2, then the opposite-sign UV-Vis TA spectra of 2D selenium accurately characterize the dynamics of absorption bleaching in this material. The latter behavior is commonly observed for other narrow bandgap 2D materials such as TI $Bi_2Se_3$ thin films (Fig. 2). As a result, the multiphoton-pumped UV-Vis TA spectra of 2D selenium converted in this way can be interpreted within the framework of generally accepted models developed for 2D materials (Section 2.2) and do not require any additional complex models, as discussed in [44].

**Conclusions**

In summary, we have reviewed experimental observations for several 2D materials in which UV-Vis TA spectra extend to energies well above the pump photon energy. This kind of spectroscopic upconversion is attributed to multiphoton (multistep) pumping in 2D materials caused by the inverse-bremsstrahlung-type free carrier absorption. This multiphoton excitation mechanism is due to collisional heating of the photoexcited electron-hole plasma within the peak intensity of the pump pulse. This behavior is due to the extreme thickness of 2D materials, which is much less than the typical depth of light penetration into their bulk counterparts and the mean free path of photoexcited carriers. It is concluded that multiphoton-pumped UV-Vis TA spectroscopy most accurately characterizes all the specific features of the ultrafast dynamics of carriers in 2D materials, since it includes their collisions with interfacial potential barriers. We also highlighted all the possible contributions to the UV-Vis TA spectra of 2D materials and considered them from the point of view of condensed matter physics. Specifically, we highlighted absorption bleaching, bandgap renormalization, and inverse-bremsstrahlung-type free carrier absorption. We also discussed the influence of valence and conduction band splitting dynamics, which typically occur in 2D materials due to their 2D nature, on the ultrafast carrier relaxation observed in multiphoton-pumped UV-Vis TA spectra. These dynamics are associated with spin splitting of the valence band and dynamic Rashba spin splitting of the conduction band induced by the built-in electric field.

**References**

1. K. Novoselov, A. Geim, S. Morozov, *et al.* Two-dimensional gas of massless Dirac fermions in graphene. Nature **438**, 197–200 (2005).
2. L. Huang, G. V. Hartland, L.-Q. Chu, R. M. Feenstra, C. Lian, K. Tahy, H. Xing, Ultrafast transient absorption microscopy studies of carrier dynamics in epitaxial graphene, Nano Letters, **10** (4), 1308-1313 (2010).
3. S. Sharma, S. Liu, J. H. Edgar, I. Chatzakis, Auger recombination kinetics of the free carriers in hexagonal boron nitride, ACS Photonics **10** (10) 3586–3593 (2023).
4. S. H. Aleithan, M. Y. Livshits, S. Khadka, J. J. Rack, M. E. Kordesch, E. Stinaff, Broadband femtosecond transient absorption spectroscopy for a CVD $MoS_2$ monolayer, Phys. Rev. B **94**, 035445 (2016).
5. E. A. A. Pogna, M. Marsili, D. D. Fazio, S. D. Conte, C. Manzoni, D. Sangalli, D. Yoon, A. Lombardo, A. C. Ferrari, A. Marini, G. Cerullo, D. Prezzi, Photo-induced bandgap renormalization governs the ultrafast response of single-layer $MoS_2$, ACS Nano, **10**, 1182–1188 (2016).
6. P. Schiettecatte, P. Geiregat, Z. Hens, Ultrafast carrier dynamics in few-layer colloidal molybdenum disulfide probed by broadband transient absorption spectroscopy, The Journal of Physical Chemistry C **123** (16), 10571-10577 (2019).
7. Q. Cui, F. Ceballos, N. Kumar, H. Zhao, Transient absorption microscopy of monolayer and bulk $WSe_2$, ACS Nano **8** (3), 2970-2976 (2014).
8. X. Hong, J. Kim, S. F. Shi, et al. Ultrafast charge transfer in atomically thin $MoS_2/WS_2$ heterostructures. Nat. Nanotech. **9**, 682–686 (2014).
9. K. Zhang, B. Jin, C. Park, Y. Cho, X. Song, X. Shi, S. Zhang, W. Kim, H. Zeng, J. H. Park, Black phosphorene as a hole extraction layer boosting solar water splitting of oxygen evolution catalysts, Nat. Commun. **10**, 2001 (2019).
10. C. Trovatello, F. Katsch, Q. Li, X. Zhu, A. Knorr, G. Cerullo, S. Dal Conte, Disentangling many-body effects in the coherent optical response of 2D semiconductors, Nano Letters, **22**, 13, 5322-5329, (2022).
11. L. Yuan, T. Wang, T. Zhu, M. Zhou, L. Huang, Exciton dynamics, transport, and annihilation in atomically thin two-dimensional semiconductors, The Journal of Physical Chemistry Letters, **8** (14), 3371-3379 (2017).
12. L. Wang, Z. Wang, H.-Y. Wang, G. Grinblat, Y.-L. Huang, D. Wang, X.-H. Ye, X.-B. Li, Q. Bao, A.T.-S. Wee, S. A. Maier, Q.-D. Chen, M.-L. Zhong, C.-W. Qiu, H.-B. Sun, Slow cooling and efficient extraction of C-exciton hot carriers in $MoS_2$ monolayer, Nat. Commun. **8**, 13906 (2017).

13. W. Wang, N. Sui, Z. Kang, Q. Zhou, L. Li, X. Chi, H. Zhang, X. He, B. Zhao, Y. Wang, Cooling and diffusion characteristics of a hot carrier in the monolayer $WS_2$, Optics Express, **29**, 7736 (2021).
14. J. C. Johannsen, S. Ulstrup, F. Cilento, A. Crepaldi, M. Zacchigna, C. Cacho, I. C. E. Turcu, E. Springate, F. Fromm, C. Raidel, T. Seyller, F. Parmigiani, M. Grioni, P. Hofmann, Direct view on the ultrafast carrier dynamics in graphene, Phys. Rev. Lett., **111**, 27403 (2013).
15. I. Gierz, J. C. Petersen, M. Mitrano, C. Cacho, I. C. E. Turcu, E. Springate, A. Stöhr, A. Köhler, U. Starke, A. Cavalleri, Nat. Mater., Snapshots of non-equilibrium Dirac carrier distributions in graphene, **12**, 1119–1124 (2013).
16. I. Gierz, S. Link, U. Starke and A. Cavalleri, Non-equilibrium Dirac carrier dynamics in graphene investigated with time- and angle-resolved photoemission spectroscopy, Faraday Discuss., **171**, 311–321 (2014).
17. S. Ulstrup, J. C. Johannsen, F. Cilento, J. A. Miwa, A. Crepaldi, M. Zacchigna, C. Cacho, R. Chapman, E. Springate, S. Mammadov, F. Fromm, C. Raidel, T. Seyller, F. Parmigiani, M. Grioni, P. D. C. King, P. Hofmann, Ultrafast dynamics of massive Dirac fermions in bilayer graphene, Phys. Rev. Lett., **112**, 257401 (2014).
18. F. Liu, Time- and angle-resolved photoemission spectroscopy (TR-ARPES) of TMDC monolayers and bilayers, Chem. Sci., **14**, 736 (2023).
19. Y. H. Wang, D. Hsieh, E. J. Sie, H. Steinberg, D. R. Gardner, Y. S. Lee, P. Jarillo-Herrero, N. Gedik, Measurement of intrinsic Dirac fermion cooling on the surface of the topological insulator $Bi_2Se_3$ using time-resolved and angle-resolved photoemission spectroscopy. Phys. Rev. Lett. **109**, 127401 (2012).
20. J. A. Sobota, S.-L. Yang, D. Leuenberger, A. F. Kemper, J. G. Analytis, I. R. Fisher, P. S. Kirchmann, T. P. Devereaux, Z.-X. Shen, Distinguishing bulk and surface electron-phonon coupling in the topological insulator $Bi_2Se_3$ using time-resolved photoemission spectroscopy. Phys. Rev. Lett. **113**, 157401 (2014).
21. J. Güdde and U. Höfer, Ultrafast dynamics of photocurrents in surface states of three-dimensional topological insulators, Phys. Status Solidi B, **258**, 2000521 (2021).
22. M. Hajlaoui, et al. Tuning a Schottky barrier in a photoexcited topological insulator with transient Dirac cone electron-hole asymmetry. Nat. Commun. **5**, 3003 (2014).
23. S. Zhu, et al. Ultrafast electron dynamics at the Dirac node of the topological insulator $Sb_2Te_3$, Sci. Rep. **5**, 13213 (2015).
24. V. Campanari, D. Catone, P. O'Keeffe, A. Paladini, S. Turchini, F. Martelli, M. Salvato, N. Loudhaief, E. Campagna, P. Castrucci, Dynamics of the bulk-to-topological state scattering of photoexcited carriers in $Bi_2Se_3$ thin films, ACS Appl. Electron. Mater., **5** (8), 4643-4649 (2023).
25. Y. D. Glinka, S. Babakiray, T. A. Johnson, M. B. Holcomb, D. Lederman, Nonlinear optical observation of coherent acoustic Dirac plasmons in thin-film topological insulators. Nat. Commun. **7**, 13054 (2016).
26. Y. D. Glinka, J. Li, T. He, X. W. Sun, Clarifying ultrafast carrier dynamics in ultrathin films of the topological insulator $Bi_2Se_3$ using transient absorption spectroscopy. ACS Photonics **8**, 1191 (2021).
27. Y. D. Glinka, T. He, X. W. Sun, Dynamic opening of a gap in Dirac surface states of the thin-film 3D topological insulator $Bi_2Se_3$ driven by the dynamic Rashba effect. J. Phys. Chem. Lett. **12**, 5593 (2021).
28. Y. D. Glinka, T. He, X. W. Sun, Two-photon IR pumped UV-Vis transient absorption spectroscopy of Dirac fermions in the topological insulator $Bi_2Se_3$. J. Phys.: Condens. Matter **34**, 465301 (2022).
29. Y. D. Glinka, T. He, X. W. Sun, Coherent surface-to-bulk vibrational coupling in the 2D topologically trivial insulator $Bi_2Se_3$ monitored by ultrafast transient absorption spectroscopy. Sci. Rep. **12**, 4722 (2022).
30. W. Lin, et al., Combining two-photon photoemission and transient absorption spectroscopy to resolve hot carrier cooling in 2D perovskite single crystals: the effect of surface layer, J. Mater. Chem. C, **10**, 16751-16760 (2022).
31. S. Ki, M. Chen, X. Liang, Optoelectronic performance characterization of $MoS_2$ photodetectors for low frequency sensing applications, J. Vac. Sci. Technol. B **39**, 062201 (2021).
32. Y. D. Glinka, S. Babakiray, T. A. Johnson, A. D. Bristow, M. B. Holcomb, D. Lederman, Ultrafast carrier dynamics in thin-films of the topological insulator $Bi_2Se_3$. Appl. Phys. Lett. **103**, 151903 (2013).
33. F. Pizzocchero, L. Gammelgaard, B. Jessen, et al. The hot pick-up technique for batch assembly of van der Waals heterostructures. Nat. Commun. **7**, 11894 (2016).
34. R. Guo, X. Bu, S. Wang, G. Zhao, Enhanced electron–phonon scattering in Janus MoSSe, New J. Phys. **21**, 113040 (2019).
35. D. Wickramaratne, F. Zahid, R. K. Lake, Electronic and thermoelectric properties of few-layer transition metal dichalcogenides, J. Chem. Phys. **140**, 124710 (2014).
36. Y. Cai, J. Lan, G. Zhang, Y.-W. Zhang, Lattice vibrational modes and phonon thermal conductivity of monolayer $MoS_2$, Phys. Rev. B **89**, 035438 (2014).
37. Y. D. Glinka, R. Cai, J. Li, T. He, X. W. Sun, Observing dynamic and static Rashba effects in a thin layer of 3D hybrid perovskite nanocrystals using transient absorption spectroscopy. AIP Adv. **10**, 105034 (2020).
38. S. K. Saini, N. K. Tailor, P. Sharma, L. Tyagi, N. Vashistha, R. Yadav, A. K. Chaudhary, S. Satapathi, M. Kumar, Revealing the substrate dependent ultrafast phonon dynamics in $Bi_2Se_3$ thin films, Adv. Mater. Interfaces, **10** (3), 2201650 (2023).
39. Q. Liu, K. Wei, Y. Tang, Z. Xu, X. Cheng, T. Jiang, Visualizing hot-carrier expansion and cascaded transport in $WS_2$ by ultrafast transient absorption microscopy, Adv. Sci., **9**, 2105746, (2022).
40. F. Ahmad, R. Kumar, S. S. Kushvaha, M. Kumar, P. Kumar, Charge transfer induced symmetry breaking in $GaN/Bi_2Se_3$ topological heterostructure device, npj 2D Materials and Applications, **6**, 12 (2022).
41. S. Kim, et al. Ultrafast carrier–lattice interactions and interlayer modulations of $Bi_2Se_3$ by x-ray free-electron laser diffraction, Nano Lett. **21**, 8554 (2021).
42. Y. Huang, et al., Ultrafast measurements of mode-specific deformation potentials of $Bi_2Te_3$ and $Bi_2Se_3$, Phys. Rev. X, **13**, 041050 (2023).
43. L. Luo, et al., Ultrafast manipulation of topologically enhanced surface transport driven by mid-infrared and terahertz pulses in $Bi_2Se_3$, Nat. Commun. **10**, 607 (2019 ).
44. S. Prodhan, K. K. Chauhan, T. Singha, M. Karmakar, N. Maity, R. Nadarajan, P. Kumbhakar, C. S. Tiwary, A. K. Singh, M. M. Shaijumon, P. K. Datta, Comprehensive excited state carrier dynamics of 2D selenium: One-photon and multi-photon absorption regimes, Appl. Phys. Lett. **123**, 021105 (2023).
45. B. S. Richards, D. Hudry, D. Busko, A. Turshatov, I. A. Howard, Photon upconversion for photovoltaics and photocatalysis: a critical review, Chem. Rev., **121** (15), 9165–9195 (2021).
46. P. Y. Yu and M. Cardona, Fundamentals of Semiconductors: Physics and Materials Properties (Springer, New York, 1996).
47. Y. D. Glinka, T. He, X. W. Sun, Characterization of charge-carrier dynamics at the $Bi_2Se_3/MgF_2$ interface by multiphoton pumped UV–Vis transient absorption spectroscopy, J. Phys.: Condens. Matter **35**, 375301 (2023).
48. V. R. Munirov and N. J. Fisch, Inverse Bremsstrahlung current drive, Phys. Rev. E **96**, 053211 (2017).
49. R. Cauble and W. Rozmus, The inverse bremsstrahlung absorption coefficient in collisional plasmas, Phys. Fluids **28**, 3387 (1985).
50. C. Kittel, Introduction to solid state physics (Wiley, 8th ed., 2004).
51. C.A. Sacchi, Laser-induced electric breakdown in water, J. Opt. Soc. Am. B **8**, 337 (1991).
52. A. Vogel, J. Noack, G. Huttman, G. Paltauf, Mechanisms of femtosecond laser nanosurgery of cells and tissue, Appl. Phys. B **81**, 1015–1047 (2005).
53. J. P. Mondia, H. W. Tan, S. Linden, H. M. van Driel, J. F. Young, Ultrafast tuning of two-dimensional planar photonic-crystal

54. waveguides via free-carrier injection and the optical Kerr effect, J. Opt. Soc. Am. **22**, 11 (2005).
54. J. G. Lu, S. Fujita, T. Kawaharamura, H. Nishinaka, Y. Kamada, T. Ohshima, Z. Z. Ye, Y. J. Zeng, Y. Z. Zhang, L. P. Zhu, H. P. He, and B. H. Zhao, Carrier concentration dependence of band gap shift in n-type ZnO:Al films, J. Appl. Phys. **101**, 083705 (2007).
55. Z. M. Gibbs, A. LaLonde, and G. J. Snyder, Optical band gap and the Burstein–Moss effect in iodine doped PbTe using diffuse reflectance infrared Fourier transform spectroscopy, New J. Phys. **15**, 075020 (2013).
56. Z. Guo, Y. Wan, M. Yang, J. Snaider, K. Zhu, and L. Huang, Long-range hot-carrier transport in hybrid perovskites visualized by ultrafast microscopy, Science **356**, 59 (2017).
57. Z. Y. Zhu, Y. C. Cheng, U. Schwingenschlögl, Giant spin-orbit-induced spin splitting in two-dimensional transition-metal dichalcogenide semiconductors, Phys. Rev. B **84**, 153402 (2011).
58. C Timm and K H Bennemann, Response theory for time-resolved second-harmonic generation and two-photon photoemission, J. Phys.: Condens. Matter **16**, 661–694 (2004).
59. H. Ohnishi and N. Tomita, Two topics of optical excitation dynamics, newly unveiled by the time- and momentum-resolved photo-electron emission from the conduction band of GaAs: A Theoretical Review, Appl. Sci., **8**, 1788 (2018).
60. M. Reutzel, A. Li, H. Petek, Above-threshold multiphoton photoemission from noble metal surfaces, Phys. Rev. B **101**, 075409 (2020).
61. K. Cheng, J. Liu, K. Cao, L. Chen, Y. Zhang, Q. Jiang, D. Feng, S. Zhang, Z. Sun, T. Jia, Ultrafast dynamics of single-pulse femtosecond laser-induced periodic ripples on the surface of a gold film, Phys. Rev. B **98**, 184106 (2018).
62. B. Rethfeld, A. Kaiser, M. Vicanek, G. Simon, Ultrafast dynamics of nonequilibrium electrons in metals under femtosecond laser irradiation, Phys. Rev. B **65**, 214303 (2002).
63. S. K. Pradhan, B. Xiao, A. K. Pradhana, Energy band alignment of high-k oxide heterostructures at $MoS_2/Al_2O_3$ and $MoS_2/ZrO_2$ interfaces, J. Appl. Phys. **120**, 125305 (2016).
64. D. Lizzit, P. Khakbaz, F. Driussi, M. Pala, D. Esseni, Ohmic behavior in metal contacts to n/p-type transition-metal dichalcogenides: Schottky versus tunneling barrier trade-off, ACS Applied Nano Materials, **6** (7), 5737-5746 (2023).
65. Q. Zhang, S. Zhang, B. A. Sperling, et al. Band offset and electron affinity of monolayer $MoSe_2$ by internal photoemission. J. Electron. Mater. **48**, 6446–6450 (2019).
66. M. Achermann, A. P. Bartko, J. A. Hollingsworth, V. I. Klimov, The effect of Auger heating on intraband carrier relaxation in semiconductor quantum rods, Nat. Phys. **2**, 557-561 (2006).
67. C. M. Cirloganu, L. A. Padilha, Q. Lin, N. S. Makarov, K. A. Velizhanin, H. Luo, I. Robel, J. M. Pietryga, V. I. Klimov, Enhanced carrier multiplication in engineered quasi-type-II quantum dots, Nat. Commun. **5**, 4148 (2014).
68. L. A. Padilha, J. T. Stewart, R. L. Sandberg, W. K. Bae, W.-K. Koh, J. M. Pietryga, V. I. Klimov, Aspect ratio dependence of Auger recombination and carrier multiplication in PbSe nanorods, Nano Letters, **13** (3), 1092-1099 (2013).
69. M. L. Mueller, X. Yan, B. Dragnea, L. S. Li, Slow hot-carrier relaxation in colloidal graphene quantum dots. Nano Lett. **11**, 56–60 (2011).
70. V. I. Klimov, Spectral and dynamical properties of multiexcitons in semiconductor nanocrystals. Annu. Rev. Phys. Chem. **58**, 635-673 (2007).
71. R. Berera, R. van Grondelle, J. T. M. Kennis, Ultrafast transient absorption spectroscopy: principles and application to photosynthetic systems, Photosynth. Res. **101**, 105 (2009).
72. К. Kaasbjerg, K. S. Bhargavi, S. S. Kubakaddi, Hot-electron cooling by acoustic and optical phonons in monolayers of $MoS_2$ and other transition metal dichalcogenides, Phys. Rev. B **90**, 165436 (2014).
73. C. Ruppert, A. Chernikov, H. M. Hill, A. F. Rigosi, T. F. Heinz, The role of electronic and phononic excitation in the optical response of monolayer $WS_2$ after ultrafast excitation. Nano. Lett. **17**, 644–651 (2017).
74. B. Miller, J. Lindlau, M. Bommert, A. Neumann, H. Yamaguchi, A. Holleitner, A. Högele, U. Wurstbauer, Tuning the Fröhlich exciton-phonon scattering in monolayer $MoS_2$, Nat. Commun. **10**, 807 (2019).
75. A. Mukherjee, D. Vasileska, A. H. Goldan, Hole transport in selenium semiconductors using density functional theory and bulk Monte Carlo, J. Appl. Phys. **124**, 235102 (2018).
76. D. Heiman, D. S. Hamilton, R. W. Hellwarth, Brillouin scattering measurements on optical glasses, Phys. Rev. B **19**, 6583 (1979).
77. J. Zizka, S. King, A. G. Every, R. Sooryakumar, Mechanical properties of low- and high-k dielectric thin films: A surface Brillouin light scattering study, J. Appl. Phys. **119,** 144102 (2016).
78. Y. D. Glinka, S. Babakiray, T. A. Johnson, M. B. Holcomb, D. Lederman, Acoustic phonon dynamics in thin-films of the topological insulator $Bi_2Se_3$, J. Appl. Phys. **117**, 165703 (2015).
79. V. I. Klimov and D. W. McBranch, Femtosecond high-sensitivity, chirp-free transient absorption spectroscopy using kilohertz lasers Opt. Lett. **23**, 277 (1998).
80. Y. D. Glinka, R. Cai, X. Gao, D. Wu, R. Chen, and X. W. Sun, Structural phase transitions and photoluminescence mechanism in a layer of 3D hybrid perovskite nanocrystals, AIP Adv. **10**, 065028 (2020).
81. R. W. Boyd, Nonlinear Optics, 3th ed. (Academic Press, Orlando, 2008).
82. Y. D. Glinka, V. Y. Degoda, S. N. Naumenko, Multiphoton mechanism of generation of elementary excitations in disperse $SiO_2$, Journal of non-crystalline solids, **152**, 219-224 (1993).
83. R. P. Chin, Y. R. Shen, V. Petrova-Koch, Photoluminescence from porous silicon by infrared multiphoton excitation, Science **270**, 776 (1995).
84. Y. D. Glinka, K. W. Lin, S. H. Lin, Multiphoton-excited luminescence from diamond nanoparticles and an evolution to emission accompanying the laser vaporization process, Appl. phys. lett., **74**, 236-238 (1999).
85. Y. D. Glinka, S. H. Lin, Y. T. Chen, Two-photon-excited luminescence and defect formation in $SiO_2$ nanoparticles induced by 6.4-eV ArF laser light, Phys. Rev. B **62**, 4733 (2000).
86. C. Hwang, D. Siegel, S. K. Mo, et al. Fermi velocity engineering in graphene by substrate modification. Sci. Rep. **2**, 590 (2012).
87. A. Chernikov, T. C. Berkelbach, H. M. Hill, A. Rigosi, Y. Li, O. B. Aslan, D. R. Reichman, M. S. Hybertsen, T. F. Heinz, Exciton binding energy and nonhydrogenic Rydberg series in monolayer $WS_2$, Phys. Rev. Lett. **113**, 076802 (2014).
88. S. Park, N. Mutz, T. Schultz, S. Blumstengel, A, Han, A. Aljarb, L.-J. Li, E. JW List-Kratochvil, P. Amsalem, N. Koch, Direct determination of monolayer $MoS_2$ and $WSe_2$ exciton binding energies on insulating and metallic substrates, 2D Mater. **5**, 025003 (2018).
89. D. Y. Qiu, F. H. da Jornada, S. G. Louie, Optical spectrum of $MoS_2$: many-body effects and diversity of exciton states. Phys. Rev. Lett. **111**, 216805 (2013).
90. Y. D. Glinka, Z. Sun, M. Erementchouk, M. N. Leuenberger, A. D. Bristow, S. T. Cundiff, A. S. Bracker, X. Li, Coherent coupling between exciton resonances governed by the disorder potential. Phys. Rev. B **88**, 075316 (2013).
91. K. F. Mak, K. He, J. Shan, T. F. Heinz, Control of valley polarization in monolayer $MoS_2$ by optical helicity, Nature Nanotech. **7**, 494 (2012).
92. H. Zeng, J. Dai, W. Yao, D. Xiao, X. Cui, Valley polarization in $MoS_2$ monolayers by optical pumping, Nature Nanotech. **7**, 490 (2012).
93. T. Cao, G. Wang, W. Han, H. Ye, C. Zhu, J. Shi, Q. Niu, P. Tan, E. Wang, B. Liu, J. Feng, Valley-selective circular dichroism of monolayer molybdenum disulphide, Nature Comm. **3**, 887 (2012).
94. A. M. Jones, H. Yu, N. J. Ghimire, S. Wu, G. Aivazian, J. S. Ross, B. Zhao, J. Yan, D. G. Mandrus, D. Xiao, W. Yao, X. Xu, Optical Generation of Excitonic Valley Coherence in Monolayer $WSe_2$, Nature Nanotech. **8**, 634 (2013).

95. K. F. Mak, C. Lee, J. Hone, J. Shan, T. F. Heinz, Atomically thin MoS$_2$: A new direct-gap semiconductor. Phys. Rev. Lett. **105**, 136805 (2010).

96. H. Zhao, S. Moehl, H. Kalt, Coherence length of excitons in a semiconductor quantum well. Phys. Rev. Lett. **89**, 097401 (2002).